\documentclass[pra,twocolumn,showpacs,superscriptaddress,groupedaddress]{revtex4-1}

\usepackage{amsfonts,hyperref}
\usepackage{graphicx}
\usepackage{amssymb}
\usepackage{epsfig}
\usepackage{amsmath}
\usepackage{xcolor}

\begin{document}

\title{Gap solitons and Bloch waves of interacting bosons in one-dimensional optical lattices:
From the weak to the strong interaction limits}

\author{T. F. Xu$^1$, X. M. Guo$^1$, X. L. Jing$^1$, W. C. Wu$^2$, and C. S. Liu$^{1,2}$}

\affiliation{$^1$College of Science, Yanshan University, Qinhuangdao 066004, China\\
$^2$Department of Physics, National Taiwan Normal University, Taipei 11677, Taiwan}

\date{\today}

\begin{abstract}
We study the gap solitons and nonlinear Bloch waves of interacting bosons
in one-dimensional optical lattices, taking into account the interaction
from the weak to the strong limits. It is shown that
composition relation between the gap solitons and nonlinear Bloch waves
exists for the whole span of the interaction strength. The linear
stability analysis indicates that the gap solitons are stable when their
energies are near the bottom of the linear Bloch band gap.
By increasing the interaction strength, the stable gap solitons can turn into
unstable. It is argued that the stable gap solitons can easily be formed in a weakly
interacting system with energies near the bottoms of
the lower-level linear Bloch band gaps.
\end{abstract}

\maketitle



\section{Introduction}

Recent development of trapping and cooling techniques has enabled
experimental realizations of Bose-Einstein condensation (BEC) of
bosonic atoms and molecules in optical
lattices \cite{RevModPhys.78.179}. When the temperature is low
and density is high, strong interaction between particles can result in
significant nonlinearity in these systems.
Associated with the periodicity and nonlinearity,
there exist two important waves in these systems, namely Bloch waves and gap solitons.
Bloch waves, which exist in both linear and nonlinear periodic systems,
are extensive and spread over the whole space \cite{Aschcoft}.
On the contrary, gap solitons, which are spatially localized atomic wave packets,
exist only in a nonlinear periodic system \cite{book}.
In particular, a class of solitons called the fundamental gap solitons (FGSs),
have the major peak well localized within a unit cell \cite{Louis}.
The solitons with two peaks of opposite signs within
a unit cell are called the subfundamental solitons \cite{PhysRevA.74.033616}.
The relationship between the nonlinear Bloch waves (NLBWs) and the
gap solitons is a topic of considerable interest.

Both FGSs and NLBWs can be simultaneously obtained by solving numerically the nonlinear
Schr\"{o}dinger equation of the system. For the weakly interacting
one-dimensional (1D) periodic Bose system, it has been shown that NLBW can be regarded as
the superposition of FGSs in an infinitely-long chain  \cite{Zhang1,Zhang2}.
This, so-called the ``composition relation", leads to the prediction
that there are $n$ families of FGSs in the $n$th band gap
of the corresponding linear periodic system.
It also implies that a class of solutions similar to the Bloch waves can be
built from it. It is interesting to see whether the
composition relation remains correct in a strongly interacting system?
To what extent the FGSs and NLBWs change when the interaction changes?

As is well known, mean-field theory typically does not work well for a 1D
system, except in the very weakly interacting regime. The
enhanced quantum fluctuation is significant in 1D quantum systems \cite
{PhysRev.130.1605, PhysRevLett.85.3745, PhysRevLett.86.5413,
PhysRevA.68.063605} which exhibit fascinating phenomena
significantly different from their three-dimensional counterparts.
On the other hand, for a single-component Bose gas
in a harmonic trap, it has been shown that with the increase
of the repulsive interaction, the
density profile evolves continuously from a
Gaussian-like distribution of bosons to a shell-structured distribution of
fermions, called the Tonks-Girardeau (TG) gas
 \cite{PhysRevLett.95.140402, PhysRevA.73.063617}. Thus when
the interaction is strong, non-perturbative methods such as
the Bose-Fermi mapping \cite{PhysRevLett.99.230402} or the Bethe ansatz \cite
{PhysRevLett.20.98, 0295-5075-61-3-368} need to be used to
characterize the features of the system properly.
Recently 1D harmonic trapped spinless Bose systems
with a repulsive $\delta$-function interaction have been
solved in the limit of $N\rightarrow\infty$ \cite{PhysRevLett.86.5413,Yang2009, Yang2010}.
Moreover, considering the two-component Bose gas with spin-independent
interactions in the absence of the external potential,
the ground-state energy density function has been extracted
from the Bethe-ansatz solution. With the local-density approximation,
a modified coupled nonlinear Schr\"{o}dinger equations for the ground state
has been obtained \cite{PhysRevA.80.043608}. Using this kind of
nonlinear Schr\"{o}dinger equation, one is able to study the interplay
between the periodicity and the nonlinearity in a 1D Bose system, especially
for the interaction from the weak to the strong limits.

In this paper, we attempt to study the NLBWs and FGSs of an interacting Bose
system in a 1D optical lattices. In particular, we are interested in
how the NLBWs and FGSs of the system change when the interaction is changed
from the weak to the strong limits, and whether the composition
relation remains correct upon the change of the interaction.
It will be shown that the composition relation remains valid in all interaction regimes.
However, linear stability analysis indicates that stability of the soliton
waves will change with the change of the nonlinearity.

The paper is organized as follows. In Sec.~\ref{Model equation},
we introduce the model equation for a 1D interacting Bose system
in optical lattices to which the interaction can be changed from
the weak to the strong limits continuously. Sec.~\ref{Ground state and
chemical potential} is devoted to compute the ground-state wave functions
and energies of the system for the
interaction from the weak to the strong limits.
It is aimed to give a direct insight of the interaction
dependent ground-state properties. By giving
as much as possible evidence numerically in Sec.~\ref{composition
relation in all interaction regime}, we show that the composition
relation exists between the NLBWs and FGSs in all interaction regimes.
A generalized composition relation between the
high-order solitons and multiple periodic waves is also shown.
In Sec.~\ref{stability of the soliton waves}, the stabilities of various soliton
waves are investigated upon the changes of the interaction and the strength
of the periodic potential. Sec.~\ref{Summary} is a brief summary.

\section{Model equation}
\label{Model equation}

We consider a 1D periodic Bose system described by the
following time-dependent nonlinear Schr\"{o}dinger equation
\begin{equation}  \label{TNSE}
i\hbar\frac{\partial \Psi(x,t)}{\partial t}=\left[ -\frac{\hbar ^{2}}{2m}\frac{%
d^{2}}{dx^{2}}+V(x) + {F}\left( \rho \right)\right] \Psi(x,t),
\end{equation}
where $V(x)=v\cos(\frac{2\pi}{\Lambda}x)$ is the periodic potential
with $\Lambda$ the lattice constant and $v$ the strength
and ${F}\left( \rho \right)$ is
responsible for the interaction energy.
The particle number is determined via $N=\int_{-\infty }^{\infty }\rho dx$ with
$\rho=|\Psi|^{2}$. To go beyond the weak interaction limit,
we use the following nonlinear form for ${F}\left( \rho \right)
$ \cite{PhysRevA.80.043608},
\begin{equation}
{F}\left( \rho \right) =\frac{\hbar ^{2}\rho ^{2}}{2m}\left[ 3e\left(
\gamma \right) -\gamma \partial _{\gamma }e\left( \gamma \right) \right],
\label{F_tilde}
\end{equation}%
where
\begin{equation}
e\left( \gamma \right) =\frac{\gamma \left( 1+p\gamma \pi ^{2}/3\right) }{%
1+q\gamma +p\gamma ^{2}}.
\label{e_gamma}
\end{equation}%
The results of Eqs.~(\ref{F_tilde}) and (\ref{e_gamma}) were recently obtained by
Hao and Chen based on Bethe Ansatz \cite{PhysRevA.80.043608}.
They can be equivalently obtained by solving
exactly the equation for $e(\gamma)$ [see,
for example, Eq.~(6) in Ref.~\cite{PhysRevLett.86.5413}],
as done, for instance in Ref.~\cite{PhysRevLett.86.5413}.
Here $\gamma \equiv {c}/{\rho}$ with $c\equiv mg/\hslash ^{2}$ ($g$ is the scattering
length) and $p=-5.0489470$ and $q=-20.8604983$ are the fitting
parameters \cite{PhysRevA.80.043608}.
In the weakly ($\gamma \ll 1$) and strongly ($\gamma \gg 1$) interacting limits,
one has the asymptotic forms
\begin{equation}
{F}\left( \rho \right) =\left\{%
\begin{array}{ll}
\displaystyle {mg\over \hbar^2}\rho \sim|\Psi|^{2}, & \gamma \ll 1 \\
\displaystyle {\pi ^{2}\hbar ^{2}\over 2m}\rho ^{2} \sim|\Psi|^{4}, & \gamma \gg 1.%
\end{array}
\right.
\label{asymptotic_F_tilde}
\end{equation}

For convenience, dimensionless scaling will be made
for the length and the energy \cite{Zhang1, Zhang2}.
Position $x$ is to be scaled in the unit of $\Lambda
/(2\pi )$;
Periodic potential $V(x)$, interaction energy ${F}(\rho)$, and
the chemical potential $\mu$ are scaled in the
unit of $8E_{r}$ with $E_{r}=\hbar ^{2}\pi ^{2}/2m\Lambda ^{2}$ being the
recoil energy and $m$ the atom mass.
Wave functions of both Bloch waves and gap solitons in optical lattices
all have the form $\Psi(x,t)=\Phi (x)\exp (-i\mu t)$.
Substituting it into the
time-dependent nonlinear Schr\"{o}dinger equation (\ref{TNSE}), one obtains
the following dimensionless time-independent nonlinear Schr\"{o}dinger equation
\begin{equation}
\left[ -\frac{1}{2}\frac{d^{2}}{dx^{2}}+v\cos(x) +F\left( \rho \right)\right]
\Phi(x) =\mu \Phi(x).  \label{reduced_chenNSE}
\end{equation}
Eq.~(\ref{reduced_chenNSE}) is the starting point of the calculations
throughout this paper.

\section{Ground-state properties}
\label{Ground state and chemical potential}

\begin{figure}[t]
\begin{center}
\includegraphics[width=9cm]{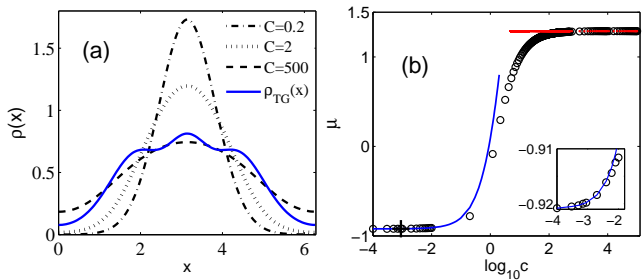}
\end{center}
\vspace{-0.5cm}
\caption{(Color online) Panel (a): The interaction constant $c$-dependent
ground-state density distribution $\rho(x)$ within one unit cell. Dimensionless
$x$ ranges from 0 to $2\pi$ for one unit length.
The large $c=500$ case is seen to behave like the
noninteracting fermions (blue line, see text).
Panel (b): The $c$-dependent ground-state
chemical potentials calculated in the
weak limit ($F(\protect\rho)\sim|\Phi |^{2}$, blue line),
in the strong limit ($F(\protect\rho)\sim|\Phi |^{4}$, red line),
and for a full $F(\protect\rho)$ (circle line)
respectively. The inset shows an enlarged view of
the curves at small $c$'s. The mark ``+", corresponding to $c=0.0009$, is
what used in Ref.~\cite{Zhang1} to discuss the nonlinear bands.}
\label{fig1}
\end{figure}

By solving Eq.~(\ref{reduced_chenNSE}) numerically, one is able
to obtain the ground-state
density profile $\rho(x)$ and the corresponding chemical potential $\mu$.
It is remarked that to solve Eq.~(\ref{reduced_chenNSE}), we
first differentiate it using the finite-element method along with the periodic
boundary condition \cite{FEM}, and then evaluate several hundreds of steps in
imaginary time until the lowest chemical potential $\mu$ is reached.
In the calculation, the system is taken to be 16 lattice constant
long ($x$ ranges from $-16\pi$ to $16\pi$),
the strength of periodic potential $v=1.5$, and the particle
number $N=48$ (average 3 particles per unit cell).
Figure~\ref{fig1}(a) shows the density profile $\rho(x)$ within a unit cell.
In view of Fig.~\ref{fig1}(a), similar to the case of the Bose gas
in a single harmonic trap,
$\rho(x)$ evolves from the Bose distribution to the Fermi-like one when
the interaction constant $c$ is increased.
In the weakly interacting regime
($c=0.2$), the density profile displays a Gaussian-like Bose distribution.
When $c$ is increased (see, for example, the case of $c=2$),
density profile decreases at the center while increases at the two sides.
In the extremely strong interaction regime ($c=500$), due to the strong
enhancement of particle tunnelling rate across neighboring wells, density profile
at the boundary increases significantly. In such limit,
so-called the TG gas, the system will behave similar to
the noninteracting fermions \cite{PhysRev.139.B500}.

In order to compare the results obtained by the imaginary-time method,
we also calculate the (exact) density profiles for the TG gas using the
Bose-Fermi mapping method.
In this case, Eq.~(\ref{reduced_chenNSE}) is first solved by the
finite-element method by setting $F(\rho)=0$. The
density profile is then calculated via $\rho_{\mathrm{TG}}(x)= \sum_{n=1}^{48}
|\Phi_n(x)|^2$ with $\Phi_n(x)$ denoting the $n$-th energy eigenstate of the
Hamiltonian. It turns out that the curve of $\rho_{\mathrm{TG}}(x)$ agrees
well with the $\rho(x)$ obtained by the imaginary-time method for the large $c=500$ case
[see Fig.~\ref{fig1}(a)]. We notice that in addition to the central peak,
$\rho_{\mathrm{TG}}(x)$ exhibits two humps at two sides near the center. 
Since the bottom of a single well behaves as an harmonic potential, the
density $\rho_{\mathrm{TG}}(x)$ obtained by us is very similar to the Tonks density
for three bosons in an harmonic potential. Three is exactly the number of particles
put on average in a single well. It has been shown that by
increasing the number of particles, the number of humps increases correspondingly 
while their amplitude decreases \cite{VMT02}.

It is well known that the dilute Bose-condensed gas can be described by the
Gross-Pitaevskii (GP) mean-field (MF) theory \cite{RevModPhys.71.463}. In fact,
MF approximation is only appropriate for a long-wavelength theory.
For short-range repulsive interactions, MF theory fails in dimension $d < 2$
to which the nonlinear term $F(\rho)\sim|\Phi|^2$ in the
GP MF theory should be replaced by $F(\rho)\sim|\Phi|^4$
from the renormalization group analysis \cite{PhysRevLett.85.1146}.
One may ask to what extent the current system under studies will approach
the above two (weak and strong) limits upon the change of
the interaction constant $c$. To uncover it, in Fig.~\ref{fig1}(b) we study
the ground-state chemical potential $\mu$ as the function of
interaction constant $c$.
In view of Fig.~\ref{fig1}(b), one sees that the curve
of $F(\rho) \sim|\Phi |^{2}$ matches well the one using full $F(\rho)$
when $c<1$. Therefore the present Bose gas
can be taken to be in the weakly interacting limit when $c<1$.
It is noted that the point $c=0.0009$, marked by ``+" sign, is what used in
Ref.~\cite{Zhang1} to discuss the nonlinear bands.
The difference between the two curves increases with the increase of
$c$. On the other side, the curve using full $F(\rho)$
matches well the one of $F(\rho) \sim|\Phi|^{4}$ when $c>4$. Therefore the present
Bose gas can be considered to be in the strong TG gas limit when $c>4$.

\section{Gap solitons and Bloch waves}
\label{composition relation in all interaction regime}

When the nonlinear interaction term $F(\rho)$ is set to zero,
Eq.~(\ref{reduced_chenNSE}) turns into the
well-known linear Mathieu equation \cite{Aschcoft}.
Its eigenfunctions are linear
Bloch waves (LBWs) and eigenvalues form the linear Bloch band (LBB). Physical
solutions are forbidden in the band gaps between neighboring bands. However,
when the nonlinear interaction exists [$F(\rho)\neq 0$],
two kinds of solutions can exist in the LBB gaps.
One is the gap soliton and another is the NLBW.
Both of them can share the same chemical potential in the LBB gaps. It has been
illustrated in Ref.~\cite{Zhang1} that for the weak limit,
$F(\rho)\sim |\Phi|^2$, the first nonlinear Bloch band (NLBB) can be
lifted into the second, third, and higher LBB gaps with
the increase of the scattering length $g$. Similarly the second NLBB can also be
lifted into the third and higher LBB gaps.
Consequently in the weak limit, NLBB in the $n$th LBB gap should
include {$n$-$1$} branches which are lifted from the lower levels.
There is only one branch of NLBB that can develop in the lowest LBB gap \cite{Zhang1}.
The present issue is to check whether the above results
remain correct in a more realistic model such that
the nonlinear interaction term is described by the full $F(\rho)$.
Our following numerical results will
provide strong evidences that the above results remain valid
for the whole span of the interaction regimes.

It is worth noting that as shown in Ref.~\cite{PhysRevLett.84.5239}, 
mean-field theory of Eq.~(\ref{TNSE}) may not correctly describe the 
physics of a 1D impenetrable Tonks gas. 
This means that in the Tonks limit, a solitonic solution may not really exist.
While this poses a limit on the solitonic solutions that we have obtained, it
is still interesting to experimentally verify our theoretical predictions.

\subsection{Interaction-dependent gap solitons and nonlinear Bloch bands}

\begin{figure}[t]
\begin{center}
\includegraphics[width=9cm]{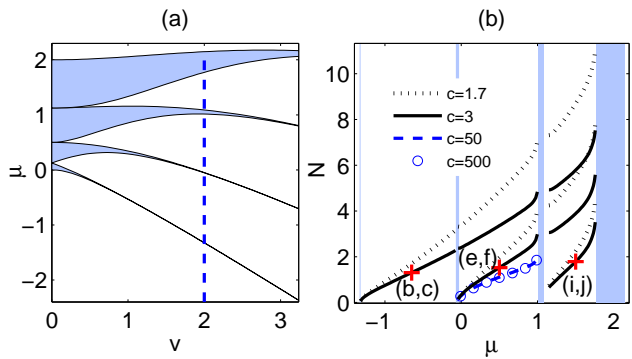}
\end{center}
\vspace{-0.5cm}
\caption{(Color online) Panel (a): Energies $\mu$ of the lowest four linear Bloch bands and
band gaps versus the strength of the periodic potential $v$.
$\mu$ and $v$ are in units of $8E_r$ with $E_r$ the recoil energy (see text).
The dashed line denotes $v=2$ used in the calculations of panel (b). Panel (b):
Particle number $N$ of FGSs as the function of chemical potential
$\protect\mu $ for different
interaction constant $c=1.7$, $3$, $50$, and $500$, respectively.
The points marked by red ``$+$"sign are those studied in
Fig.~\protect\ref{fig4}.}
\label{fig2}
\end{figure}

We first set $F(\rho )=0$ to solve the linear Schr\"{o}dinger equation (\ref
{reduced_chenNSE}) exactly by the finite-element method. The periodic
potential strength $v$-dependent
lowest four bands and band gaps are shown in Fig.~\ref{fig2}(a).
As shown, with the increase of $v$,
the continuum energy spectrum turns into energy bands with gaps.
When $v$ is large enough, some (lower) bands reduce to highly-degenerate thin
levels. We next retain $F(\rho)$ to solve the nonlinear Schr\"{o}dinger
equation (\ref{reduced_chenNSE}) numerically. In particular,
we are interested in the solutions of FGSs (or NLBWs) with the energy $\mu$
falling into the LBB gap regions. The numerical solutions are
obtained by differentiating Eq.~(\ref%
{reduced_chenNSE}) on a finite difference grid to obtain a coupled algebraic
equations and solve it with the Newton-relaxation method \cite%
{PhysRevA.74.033616}.

Figure \ref{fig2}(b) shows particle number $N$ of gap solitons
as the function of $\mu$ for
four different interaction constant $c$'s. As shown,
for example, when $c=3$ close to the strong limit (solid lines),
the first NLBB develops from the first LBB and is lifted with the
increase of $N$. The so-called NLBB lifting is simply due to the fact that
the larger $N$ is, the larger the nonlinearity and hence
the corresponding $\mu$ are. The first NLBB enters the second
LBB gap at $N\simeq 2.38$ and when $N\simeq 4.92$ it further enters the third LBB gap.
For the second NLBB, it develops from the second LBB at $N\simeq 0.14$
and is lifted in a similar manner with the increase of $N$.
It enters the third LBB gap when it crosses the critical value of $N\simeq 3.06$.
It turns out that the third NLBB only develops at $N\agt 0.66$,
slightly above the third LBB.

For relatively weaker interaction case, $c=1.7$ (dotted lines),
the NLBB is seen to correspond to larger $N$ for the same $\mu$,
as compared to those of the $c=3$ case. This is because for the smaller
interaction constant $c$, one needs a larger $N$ to
achieve the same nonlinear effect. We have also calculated the extremely
strong interaction cases (for $c=50$ and $500$) for energies within the second LBB gap.
The lines of the two cases are seen to coincide and indicate that the current
Bose system is reduced to the TG gas when the interaction constant $c$ is large enough.

\subsection{Interaction-dependent amplitude of gap solitons}

\begin{figure}[t]
\begin{center}
\includegraphics[width=7cm]{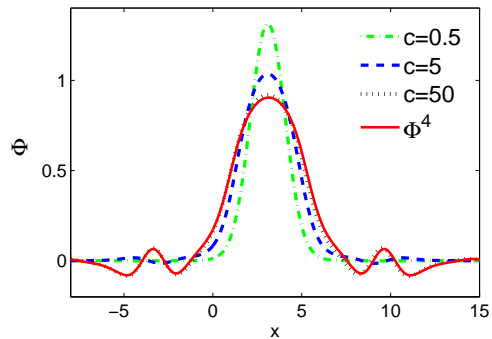}
\end{center}
\vspace{-0.5cm}
\caption{(Color online) Amplitudes of the first-family FGS as the function of the
interaction constant $c$. For all curves, particle number is fixed at $N=3$ and
periodic potential strength is $v=2$. For comparison, the case of
the strong interaction limit, $F(\rho)\sim |\Phi|^4$, is also shown.}
\label{fig3}
\end{figure}

The soliton wave occurs only in the nonlinear system and has no classical
correspondence. One may guess that the amplitude of soliton wave increases
with the increase of the nonlinearity. However, the truth is in the
opposite direction. As shown in Fig.~\ref{fig3},
numerical results of the first-family FGS waves show that their
amplitudes actually decrease as $c$ increases.
For all curves in Fig.~\ref{fig3}, particle number is fixed at $N=3$ and
periodic potential strength is $v=2$. For comparison,
we have also shown the case for
the strong interaction limit [$F(\rho)\sim |\Phi|^4$].


In fact, FGSs originate from the nonlinearity and periodicity of the system. On
one hand, FGSs can be taken roughly as the discrete eigenstates of the system in a
single potential well. On the other hand, FGSs are eigenstates of the
Schr\"{o}dinger equation (\ref{reduced_chenNSE}) if $[V(x)+F(\rho)]$ is taken
approximately as the effective field of a single well. When the
interaction constant $c$ is increased and for the particle number $N$ fixed, the
confinement due to the effective single well will become
weaker. This means that the wave function will become more dispersed and
correspondingly the amplitude of the FGSs will become smaller.

\subsection{Composition relation at strong interactions}

As shown previously, NLBB can be viewed as the lifted LBB by
increasing the nonlinear interaction. While LBB can be viewed as the
evolution from the discrete energy levels of an individual well.
Therefore it is not difficult to see
that NLBW belonging to the $n$th NLBB should have {$n$-$1$} nodes (in the sense of
the $n$th bound state) in an individual well of the periodic potential.
It has been pointed out and proven numerically for the weak
interaction limit [$F(\rho)\sim|\Phi|^2$] that FGSs which can only
exist in a nonlinear system
are the building blocks of the NLBWs \cite{Zhang1}.
Thus it is expected that FGS should behave like the bound states in an
individual well of the periodic potential. The current issue is to
verify whether the above conclusion remains correct when the nonlinear
term is replaced by the full $F(\rho)$.
\begin{figure}[t]
\begin{center}
\includegraphics[width=8.0cm]{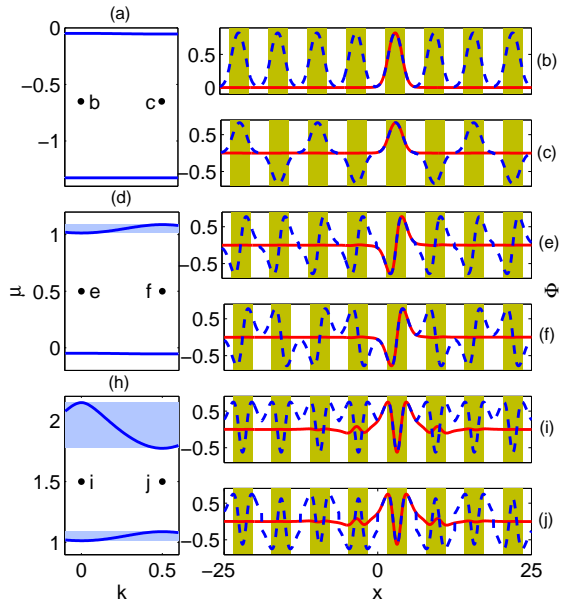}
\end{center}
\vspace{-1cm}
\caption{(Color online) Left column: panel (a),(d), and (h) show the first,
second, and third LBBs and band gaps. Thick black dots denote the
$(k,\mu)$ points studied in the right column.
Right column: panel (b) and (c) are the FGS (red solid line) and
NLBW (blue dashed line) in the first LBB gap with $N=1.30$ and
$(k,\mu)=(0,-0.65)$ (at BZ center) and $(0.5,-0.65)$ (at BZ edge)
respectively.
Panel (e) and (f) are the FGS and NLBW in the second LBB gap with $N=1.52$
and $(k,\mu)=(0,0.5)$ and $(0.5,0.5)$ respectively; panel (i) and (j) are
the FGS and NLBW in the third LBB gap with $N=1.79$ and $(k,\mu)=(0,1.5)$
and $(0.5,1.5)$ respectively. For all panels, the
interaction constant is $c=3$ and the periodic potential strength is $v=2$.
The $(N,\mu)$ points under study are also marked in Fig.~\protect\ref{fig2}(b).}
\label{fig4}
\end{figure}

In Fig.~\ref{fig4}, we solve and plot both NLBW (blue dashed line)
and FGS (red solid line) for the relatively strong interaction ($c=3$) case.
It should be emphasized that the conclusion drawn here remains
valid for much more stronger interaction cases.
We consider both NLBWs and FGSs in the first, second, and third
LBB gap respectively and for the representative we take the $k$ points
at both the BZ center ($k=0$) and boundary ($k=0.5$).
In view of Fig.~\ref{fig4}(b)\&(c) with $\mu=-0.65$ and $N=1.30$,
an almost perfect match is found between the NLBW and the
FGS within one unit cell in the first LBB gap.
The good match occurs regardless of
at the center or at the edge of the BZ.
It thus gives a strong evidence that FGSs can be
considered as the building blocks of the NLBWs. It is important to note that
the waves shown in Fig.~\ref{fig4}(b)\&(c) are belonging to the first-family
FGS and hence of no node. In Fig.~\ref{fig4}(e)\&(f),
a good match is also found between the NLBW and the FGS within one unit cell
in the second LBB gap.
These waves with $\mu=0.5$ and $N=1.52$ are belonging to the second-family
FGS and hence of one node.
There is a key difference between the NLBWs in Fig.~\ref{fig4}(b) and (c)
such that the waves are oscillating in-phase ($k=0$) between the neighboring
wells in case (b), while it is out-of-phase ($k=0.5$) in case (c).
Similar consequence is also seen between Fig.~\ref{fig4}(e) and (f) and
also between Fig.~\ref{fig4}(i) and (j).


As a test, we actually try to build the possible NLBWs using
the first obtained third-family FGS in the third LBB gap [see
Fig.~\ref{fig4}(i) and (j)]. These third-family
FGSs with $\mu=1.5$ and $N=1.79$ are obtained by the Newton-relaxation
method with proper initial condition.
To capture the out-of-phase motion of the NLBWs
at the BZ edge [case (j)], we have added
an alternative negative sign to the building block.
In view of Fig.~\ref{fig4}(i) and (j),
the match between the NLBW and FGS is good except
FGS exhibits a short tail across the cell boundary. Nevertheless, one
can still conclude that the composition relation between NLBWs
and FGSs is valid to a good approximation.

\subsection{Generalized composition relation}

In addition to the FGSs and NLBWs studied previously,
there are two types of waves which are also common
in a nonlinear system. One is called the high-order gap solitons
which are gap soliton waves of multiple peaks over multiple unit cells \cite{Wang}.
As we will show later, high-order gap solitons can be viewed as the truncated NLBW
and hence NLBWs can be built from them. Another is called
the multiple periodic waves which are defined as $%
\Phi(x)=\exp(ikx)\psi_{k}(x)$ with $\psi_{k}(x)=\psi_{k}(x+2n\pi)$ and $n$
being a positive integer \cite{PhysRevA.69.043604}.
In the weakly interacting limit, it has been shown that
composition relation between the FGSs and the NLBWs can be
generalized to construct multiple periodic waves
from the high-order gap solitons \cite{Zhang2}. The present issue
is again to see whether the generalized composition relation
remains valid in the strong interaction limit.

\begin{figure}[t]
\begin{center}
\vspace{-1cm}
\includegraphics[width=8cm]{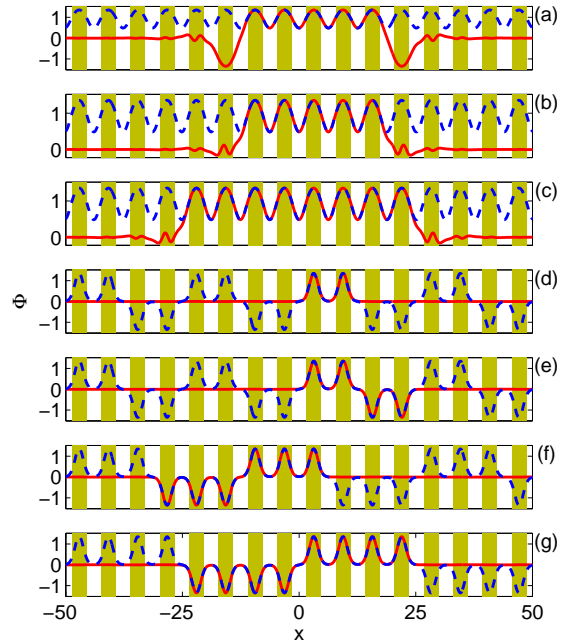}
\end{center}
\vspace{-0.3cm}
\caption{(Color online) Panel (a)--(c):
Illustration of the high-order gap solitons (red solid line) and NLBWs
(blue dashed line) for $c=3$ and $\protect\mu =1.63$.
(a) The upper power branch of five-peak gap solitons;
(b) The lower power branch of five-peak gap solitons;
(c) The lower power branch of eight-peak gap solitons.
Panel (d)--(g): Illustration of the high-order gap solitons
(red solid line) and multiple periodic waves (blue dashed line)
for $c=0.3$ and $\protect\mu =-1.07$. (d) Two-peak solitons with $N=6$;
(e) Solitons of two peaks up and two peaks down with
$N=12$; (f) Solutions of three peaks up and three peaks down with $N=18$; (g)
Solitons of four peaks up and four peaks down with $N=24$. The periodic potential strength is $v=2$ for all panels.}
\label{fig5}
\end{figure}

We have first solved the time-dependent
nonlinear Schr\"{o}dinger equation (\ref{TNSE})
numerically using the imaginary-time method to
obtain NLBW function $\Phi(x)$ and chemical potential $\mu$
with a given interaction constant $c$. The Newton-relaxation method is then
used to solve the time-independent
nonlinear Schr\"{o}dinger equation (\ref{reduced_chenNSE})
with the pre-obtained $\mu$ and the proper initial condition.
Fig.~\ref{fig5}(a)--(c) show the five- and eight-peak high-order gap
solitons with the interaction constant $c=3$. An
almost perfect match is found between the high-order gap solitons and
the NLBWs. Thus high-order gap solitons can also be viewed as
the truncated NLBWs for the strong interactions.

Fig.~\ref{fig5}(d)-(g) show the composition relation between the
high-order gap solitons and multiple periodic waves. Here $c=0.3$ is used.
Shown in Fig.~\ref{fig5}(d) is the building block of two-peak high-order
gap solitons which generates the two-peak up and two-peak down
multiple periodic wave.
Fig.~\ref{fig5}(e) shows the same two-peak up and two-peak down
multiple periodic wave that can also be built by the two-peak up and two-peak
down high-order gap solitons. We further show the three-peak up
and three-peak down and four-peak up
and four-peak down multiple periodic waves in Fig.~\ref{fig5}(f) and (g)
built by the three-peak up and three-peak down and four-peak up
and four-peak down high-order gap solitons respectively.
The matches between the higher-order gap solitons and
multiple periodic waves are all seen to be good.
This stands that for the whole
range of the interactions, FGSs are the most basic building blocks which
can be used to build high-order solitons and multiple
periodic waves.

\section{stability of the soliton waves}

\label{stability of the soliton waves}

Stability of the FGS is another important issue when the interaction
of the system is changed from the weak to the strong limit.
We shall study the linear stability of FGS solution following the
standard procedure. Since the unstable solution is sensitive to a small
perturbation, one can add a small perturbation $\Delta \Phi (x,t)$ to a
known solution $\Phi(x)$
\begin{equation*}
\Psi (x,t)=\left[ \Phi (x)+\Delta \Phi (x,t)\right] \exp (-i\mu t),
\label{perturb}
\end{equation*}
where $\Delta \Phi (x,t)=u(x)\exp (i\alpha t)+w^{\ast }(x)\exp (-i\alpha
^{\ast }t)$.
Inserting the perturbation into Eq.~(\ref{TNSE}) and dropping the higher-order terms
in ($u,v$), one then obtains the linear eigenequation
\begin{equation}
\left(
\begin{array}{cc}
\mathcal{L} & -\frac{\partial F\left( \rho \right) }{\partial \rho }\Phi ^{2}
\\
\frac{\partial F\left( \rho \right) }{\partial \rho }\Phi ^{\ast 2} & -%
\mathcal{L}%
\end{array}%
\right) \left(
\begin{array}{c}
u \\
w%
\end{array}%
\right) =\alpha \left(
\begin{array}{c}
u \\
w%
\end{array}%
\right),  \label{LinearStability}
\end{equation}%
where $\mathcal{L}\equiv\frac{1}{2}\frac{d^{2}}{dx^{2}}-V\left( x\right) -F\left(
\rho \right) -\frac{\partial F\left( \rho \right) }{\partial \rho }%
\left\vert \Phi \left( x\right) \right\vert ^{2}+\mu. $ Linear stability of
a soliton is determined by the energy spectrum of the linear eigenequation (\ref%
{LinearStability}). Among all eigenvalues $\alpha$ obtained,
if there exists one of a finite
imaginary part, the solution of $\Phi (x)$ would be unstable.
Otherwise, the solution of $\Phi (x)$ is stable.

\begin{figure}[t]
\begin{center}
\includegraphics[width=8cm]{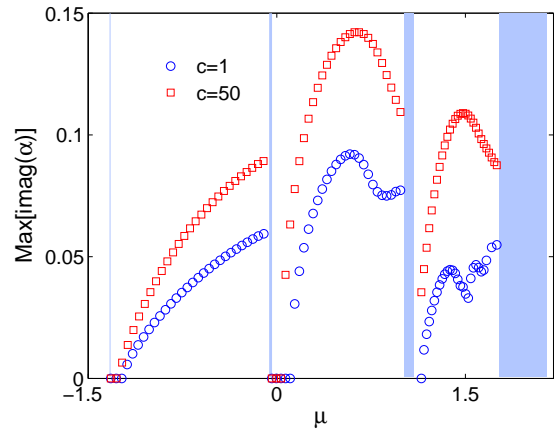}
\end{center}
\caption{(Color online) Studies of the stability of the FGSs in the
first, second, and third LBB gap for weak ($c=1$) and strong ($c=50$)
interaction limits. The periodic potential strength used is $v=2$.}
\label{fig6}
\end{figure}

The stability of FGSs is investigated in Fig.~\ref{fig6}. For the weak
interaction case ($c=1$), the linear stability analysis indicates that the first
family FGSs which develop from the first LBB are stable when
their chemical potentials $\mu$ are
near the bottom of the first LBB gap. They will become unstable
when $\mu$ becomes higher within the first band gap (with the increase of $N$), and
enters into the second and third band gaps (not shown in Fig.~\ref{fig6}).
Similarly for the second family FGSs
which develop from the second LBB, they are also stable when
the chemical potentials are
near the bottom of the second band gap. While for the third family FGSs,
they are stable only when $\mu$ is extremely close to the
bottom of the third band gap. In contract, for the strong interaction
case ($c=50$), studies of linear stability indicate that
stable FGSs of the first and second families exist only in a much narrower
regime with $\mu$ near the bottom of the band gaps;
while it seems that the third-family FGSs can no longer be stable.
Bearing the above results, it is argued that stable FGSs can be easily formed
in a weakly interacting system with chemical potential
near the bottom of the lower-level band gaps.

\begin{figure}[t]
\begin{center}
\includegraphics[width=8cm]{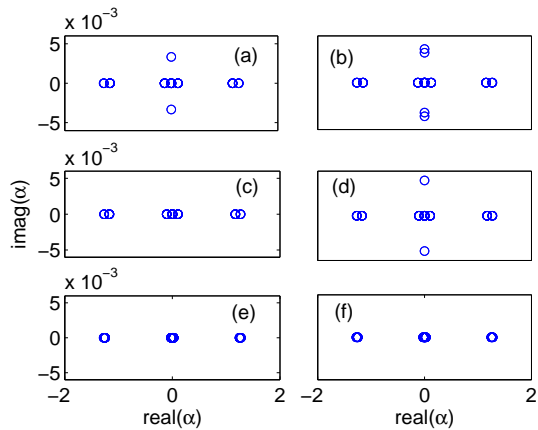}
\end{center}
\caption{(Color online) Linear-stability studies of the five-peak gap solitons in the
first LBB gap. Panel (a), (c), and (e) correspond to the lower power ones shown
in Fig.~\ref{fig5}(b), while panel (b), (d), and (f) correspond to the
upper power ones shown in Fig.~\ref{fig5}(a). Chemical potential
$\protect\mu=-1.3$ for (e) and (f), $-1.22$
for (c) and (d), and $-1.2$ for (a) and (b). Interaction constant $c=1$
and periodic potential strength is $v=2$.}
\label{fig7}
\end{figure}

For completeness, we have also investigated the
stability of two kinds of five-peak gap solitons.
One is the upper power branch shown in
Fig.~\ref{fig5}(a) and another is the lower power one shown in
Fig.~\ref{fig5}(b) \cite{Wang}. The linear stability analysis
was carried out in the first LBB gap for three chemical potentials, $\mu=-1.3$, $
-1.22$, and $-1.2$ respectively and
the results are shown in Fig.~\ref{fig7}. For the case of $\mu=-1.3$,
five-peak gap solitons are stable for both upper [see Fig.~\ref
{fig7}(f)] and lower power branches [see Fig.~\ref{fig7}(e)]. This is consistent
with the case of FGSs which are stable with $\mu$ near the bottom of the LBB gap.
For the case of $\mu=-1.22$ near the middle
of the first band gap, the lower power branch of five-peak gap solitons
are still stable [see Fig.~\ref{fig7}(c)], while the upper power branch of
five-peak gap solitons become unstable [see Fig.~\ref{fig7}(d)]. This is
consistent with the one analysis studied in Ref.~\cite{Wang}. When
$\mu$ is increased to be $-1.2$, five-peak gap solitons are unstable for both the
lower [see Fig.~\ref{fig7}(a)] and upper power branches [see Fig.~\ref
{fig7}(b)]. Therefore, the stable multi-peak solitons can also be easily formed
with $\mu$ near the bottom of the lower band gaps.

The stability of the multi-peak solitons upon the change of the
periodic potential strength ($v$) is also studied. It is found that
all the unstable multi-peak solitons studied in Fig.~\ref{fig7} will eventually
become stable when $v$ is increased from the present value $2$ to $10$.
The reasons behind it are easily understood. When $v$
is increased, interaction is reduced relatively.
Consequently chemical potential will approach relatively closer to
the bottom of the LLB gap and gap solitons will become relatively more stable.

\section{Summary}

\label{Summary}

In summary, we have investigated the composition relation between the
fundamental gap solitons (FGSs) and nonlinear Bloch waves (NLBWs) of interacting
bosons in one-dimensional optical lattices. The main focus is to consider
the interaction from the weak to the strong limits.
Our numerical results verify that the composition
relation remains correct in the whole span of the interaction strength.
FGSs are thus the fundamental building blocks to build all other
stationary solutions, including NLBWs, high-order solitons, and multiple
periodic waves, in one-dimensional nonlinear periodic systems.
By the linear stability analysis, it is found that the
stable FGSs exist near the bottom of the linear band gap.
By increasing the interaction strength, the stable gap solitons will
become unstable. However, one can restore the stability of the unstable
FGSs by increasing the strength of the periodic potential.
It is argued that the stable gap solitons can be easily formed
in a weakly interacting system with the chemical potential
near the bottom of the lower-level band gaps.

\begin{acknowledgments}
We thank Yong-Ping Zhang, Biao Wu, and Shu Chen for useful discussions.
This work is supported by National Science Council of Taiwan (Grant No.
99-2112-M-003-006), Hebei Provincial Natural Science Foundation of China (Grant
No. A2010001116), and the National Natural Science
Foundation of China (Grant No. 10974169). We also acknowledge the
support from the National Center for Theoretical Sciences, Taiwan.
\end{acknowledgments}


%

\end{document}